\documentclass{Interspeech}

\usepackage{times}
\usepackage{latexsym}
\usepackage{booktabs}
\usepackage{verbatim}
\usepackage[table,xcdraw]{xcolor}
\usepackage{multirow}
\usepackage{colortbl}  
\usepackage{booktabs}

\usepackage{etoolbox}
\makeatletter
\patchcmd{\affiliation}
  {\renewcommand{\affiliationsep}{\vskip1pt}}
  {\renewcommand{\affiliationsep}{, }}
  {}{}
\makeatother

\usepackage[T1]{fontenc}
\usepackage[utf8]{inputenc}
\usepackage{microtype}
\definecolor{lightred}{rgb}{1.0, 0.8, 0.8}
\definecolor{lightblue}{rgb}{0.8, 0.9, 1.0}
\definecolor{lightgreen}{rgb}{0.8, 1.0, 0.8}
\definecolor{lightyellow}{rgb}{1.0, 1.0, 0.8}
\definecolor{lightpurple}{rgb}{0.9, 0.8, 1.0}
\definecolor{lightorange}{rgb}{1.0, 0.9, 0.8}
\usepackage{inconsolata}
\usepackage{adjustbox}
\usepackage{ subcaption, algorithm2e, url, float, etoolbox, adjustbox, pgf,booktabs,verbatim}
\usepackage{xcolor}
\usepackage{amsmath}
\usepackage{amssymb}
\usepackage{tcolorbox}
\usepackage{amsfonts}
\definecolor{litepurple5}{RGB}{237,231,246}   % Lightest
\definecolor{litepurple10}{RGB}{209,196,233}
\definecolor{litepurple15}{RGB}{179,157,219}
\definecolor{litepurple20}{RGB}{149,117,205}
\definecolor{litepurple25}{RGB}{126,87,194}    % Darkest
\usepackage[table]{xcolor}
\definecolor{strawberry5}{RGB}{255,230,230}   % Light strawberry
\definecolor{strawberry10}{RGB}{255,204,204}  % Medium strawberry
\definecolor{strawberry15}{RGB}{255,153,153}  % Dark strawberry

\title{\textbf{\texttt{HYFuse}}: Aligning Heterogeneous Speech Pre-Trained Representations in Hyperbolic Space for Speech Emotion Recognition}
\author[affiliation={1}]{Orchid Chetia}{Phukan*}
\author[affiliation={1,2}]{Girish*}{}
\author[affiliation={1,3}]{Mohd Mujtaba}{Akhtar*}
\author[affiliation={4}]{Swarup Ranjan}{Behera}
\author[affiliation={5}]{Pailla Balakrishna}{Reddy}
\author[affiliation={1}]{Arun Balaji}{Buduru}
\author[affiliation={6,7}]{Rajesh}{Sharma}

\affiliation{}{IIIT-Delhi}{India}
\affiliation{}{UPES}{India}
\affiliation{}{V.B.S.P.U}{India}
\affiliation{}{Independent Researcher}{India}
\affiliation{}{Reliance AI}{India}
\affiliation{}{University of Tartu}{Estonia}
\affiliation{}{Plaksha University}{India}

\email{\textcolor{blue}{\texttt{Correspondence:}} orchidp@iiitd.ac.in} 

\usepackage{comment}

\begin{document}

\maketitle
% \maketitle
\begingroup
  % switch to symbol‐style so \footnotetext uses “*” for the first footnote
  \renewcommand{\thefootnote}{\fnsymbol{footnote}}
  \setcounter{footnote}{0}
   \footnotetext{* Contributed equally as a first authors.}
\endgroup
\begin{abstract}
\noindent Compression-based representations (CBRs) from neural audio codecs such as EnCodec capture intricate acoustic features like pitch and timbre, while representation-learning-based representations (RLRs) from pre-trained models trained for speech representation learning such as WavLM encode high-level semantic and prosodic information. Previous research on Speech Emotion Recognition (SER) has explored both, however, fusion of CBRs and RLRs haven't been explored yet. In this study, we solve this gap and investigate the fusion of RLRs and CBRs and hypothesize they will be more effective by providing complementary information. To this end, we propose, \textbf{\texttt{HYFuse}}, a novel framework that fuses the representations by transforming them to hyperbolic space. With \textbf{\texttt{HYFuse}}, through fusion of x-vector (RLR) and Soundstream (CBR), we achieve the top performance in comparison to individual representations as well as the homogeneous fusion of RLRs and CBRs and report SOTA. 
\end{abstract}

\section{Introduction}
Speech Emotion Recognition (SER) plays a pivotal role in human-computer interaction, enabling systems to identify and understand the emotional nuances expressed in speech~\cite{Tahon2016TowardsAS}. By analyzing vocal attributes such as pitch, tone, and rhythm, SER systems uncover the intricate nuances of human affect. SER has far-reaching implications, from enhancing mental health monitoring in healthcare~\cite{Souganciouglu2020IsEF} to transforming educational tools by gauging student engagement~\cite{Tanko2022ShoelacePS}. %Despite significant advancements, the full potential of SER remains unrealized, urging continued innovation to improve its accuracy, efficiency, and applicability across diverse domains. 
Initial research in SER mostly focused on the usage of spectral features such MFCC with classical ML models such as GMM \cite{6514336}, SVM \cite{milton2013svm}. This was succeeded by the use of deep learning models such as LSTM \cite{wang2020speech}, CNN, CNN-LSTM \cite{hazra2022emotion}, etc. \par
Recent strides in SER research have seen the usage of representations from state-of-the-art (SOTA) Pre-trained models (PTMs). These representations have provided substantial performance benefit and has led to sufficient development in SER. These representations can be primarily categorized into two types: Representation-learning based representations (RLRs) derived from speech PTMs such as Wav2vec2\cite{baevski2020wav2vec}, WavLM \cite{chen2022wavlm}, XLS-R \cite{babu22_interspeech}, etc. and compression-based representations (CBRs) extracted from neural audio codecs such as EnCodec~\cite{defossez2022high}, DAC~\cite{kumar2024high}, and Soundstream~\cite{zeghidour2021soundstream}. PTMs for RLRs are generally trained for speech representation learning and it can be both for a particular language or multilingual, however, neural audio codecs (NACs) are trained for compression of input data following a encoder-decoder modeling architecture. Researchers have explored various RLRs such as Wav2vec2 \cite{pepino21_interspeech}, HuBERT \cite{morais2022speech}, etc. for SER. Also, usage of compression-based representations (CBRs) from NACs for SER has gained recent traction in the community. Wu et al. \cite{wu-etal-2024-codec} gave a initial exploration of CBRs for SER by investigating different NACs such as Encodec, DAC, Speech Tokenizer and so on. However, they only focused on English SER. Ren et al. \cite{ren2024emo} extended to chinese SER and gave a much more comprehensive analysis of various SOTA CBRs with the inclusion of more NACs. Furthermore, Mousavi et al. \cite{mousavi2024dasb} presented the first comparative study of CBRs and RLRs for SER. Further. Wu et al. \cite{10317508} also explored the fusion of RLRs such as Wav2vec2, WavLM, Unispeech-SAT, and so on for more due to existence of complementary behavior of such representations for more improved SER. Such improvement due to the combination of PTM representations can also be seen across various related speech processing tasks such as speech recognition \cite{arunkumar22b_interspeech}, synthetic speech detection \cite{combei2024wavlm}.\par
However, no focus on the fusion of heterogeneous representations i.e. RLRs and CBRs have been given, despite extensive research into SER with PTM representations. In this work, for the first time, to the best of our knowledge, we explore such fusion of heterogenenous representations (RLRs and CBRs). \textit{We hypothesize that fusion of RLRs and CBRs will lead to further improvement in SER performance by the exploitation of complementary information of RLRs and CBRs. CBRs captures the low-level features like pitch, timbre and RLRs encodes higher-level prosodic patterns.} To aid in effective fusion, we propose a novel framework, \textbf{\texttt{HYFuse}} (Fusion in Hyperbolic Space) that transforms the representations from euclidean space to hyperbolic space and performs fusion through mobius addition. As far as we know, this is the first study to investigate the usage of hyperbolic space for fusion of representations in the context of SER. The fusion of CBRs and RLRs in hyperbolic space allows for the preservation of their hierarchical relationships and complementary features, ensuring that both low-level acoustic details and high-level prosodic patterns are effectively aligned and integrated. %This will enable more accurate and robust SER by capturing the nuanced interactions between these distinct representations.

\begin{figure*}[hbt!]
    \centering
    \includegraphics[width=0.7\linewidth]{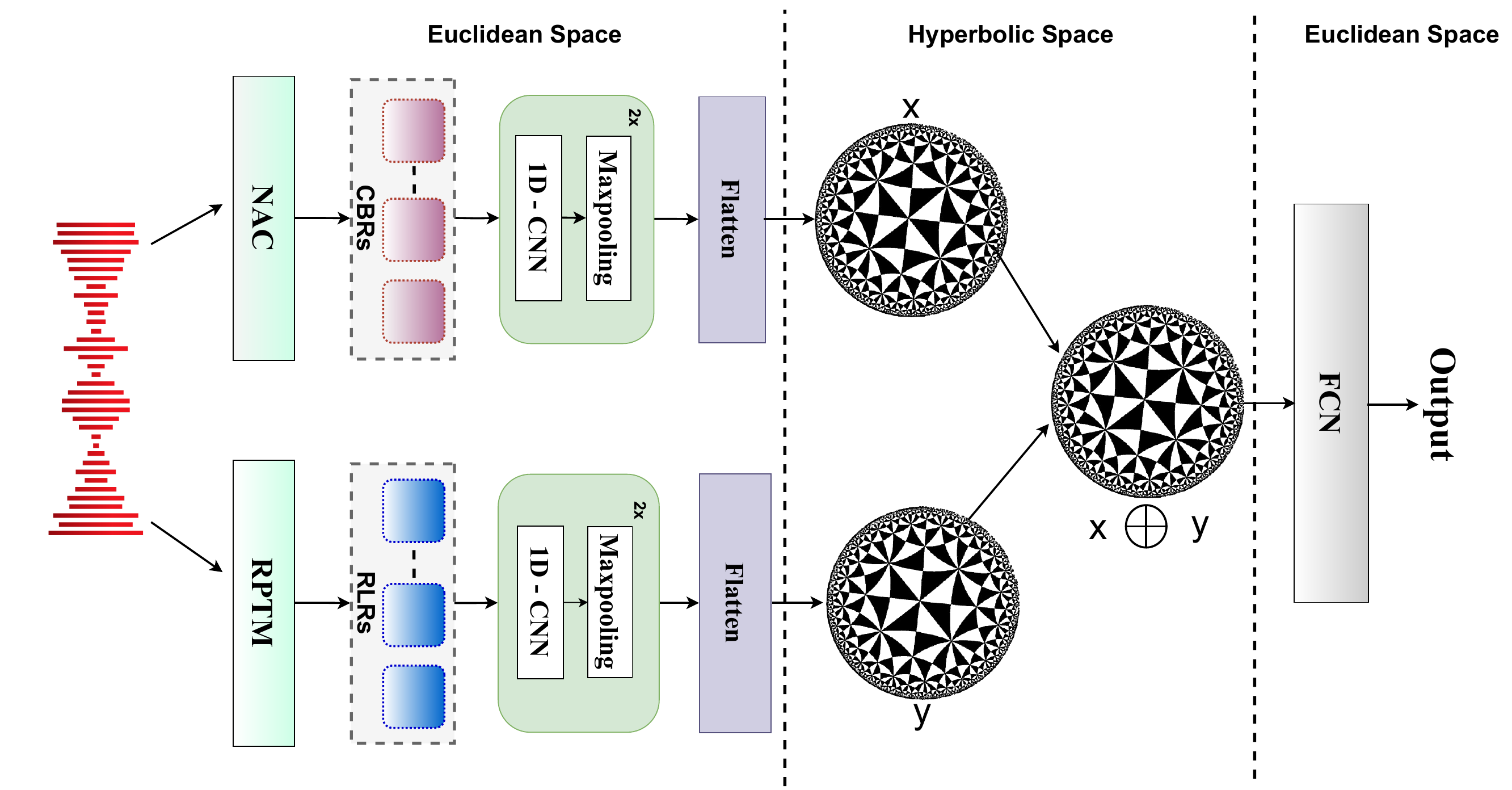}
    \caption{\textbf{\texttt{HYFuse}}; $x \oplus y$ represent mobius addition}
    \label{archi}
\end{figure*}

\noindent \textbf{The key contributions of this work are:}
\begin{itemize}
    \item We propose, \textbf{\texttt{HYFuse}} (Figure \ref{archi}), a novel framework for fusing RLRs and CBRs by transforming them into hyperbolic space and leveraging the strengths of hyperbolic geometry for effective fusion of RLRs and CBRs. 
    \item Using \textbf{\texttt{HYFuse}}, with the fusion of x-vector (RLR) and SoundStream (CBR) representations, we achieve superior performance compared to individual representations and homogeneous fusions of RLRs and CBRs. Our framework sets new SOTA results on the CREMA-D and Emo-DB benchmark datasets, establishing the efficacy of combining RLRs and CBRs for SER.
\end{itemize}
\noindent We have released the code and models from this work at: \url{https://github.com/Helix-IIIT-Delhi/HYFuse-SER}

% We will open-source the codes and model after the double-blind review process. 

% \subsection{Try}

\section{Pre-Trained Representations}
In this section, we give a brief overview of the PTMs and NACs behind RLRs and CBRs.

\subsection{Representation-learning}
WavLM\footnote{\url{https://huggingface.co/microsoft/wavlm-base}} \cite{chen2022wavlm} has shown SOTA performance in multiple speech processing tasks within SUPERB. We adopt its base version, which consists of 94.70 million parameters and is pre-trained on 960 hours of Librispeech. Similarly, Wav2Vec2\footnote{\url{https://huggingface.co/facebook/wav2vec2-base}} \cite{baevski2020wav2vec} is included in our study as a contrastive learning-based representation-learning based PTM. We use its base variant, which has 95.04 million parameters and is also pre-trained on 960 hours of Librispeech. We also incorporate HuBERT\footnote{\url{https://huggingface.co/facebook/hubert-base-ls960}} \cite{hsu2021hubert}, a speech PTM inspired by the BERT architecture, pre-trained on 960 hours of Librispeech. We utilize its base version, which consists of 94.68 million parameters. Additionally, we consider x-vector\footnote{\url{https://huggingface.co/speechbrain/spkrec-xvect-voxceleb}} \cite{snyder2018x}, a time delay neural network designed specifically for speaker recognition, comprising 4.2 million parameters. It is particularly relevant to our research, as its representations have been shown to be effective for SER \cite{chetiaphukan23_interspeech}. All audio recordings are resampled to 16 kHz before passing to the PTMs. The PTMs remain frozen, and we extract RLRs from their final hidden states using average pooling. The resulting feature dimensions are 768 for WavLM, Wav2Vec2, and HuBERT, while x-vector produces 512-dimensional representations. \par

\subsection{Compression}
Soundstream\footnote{\url{https://github.com/haydenshively/SoundStream}} \cite{zeghidour2021soundstream} is an efficient NAC designed for low-bitrate compression, utilizing an encoder-decoder architecture with Residual Vector Quantization (RVQ) and multi-scale STFT discriminators to maintain a balance between compression and audio quality. It supports bitrates ranging from 3 kbps to 18 kbps. Descript Audio Codec  (DAC)\footnote{\url{https://huggingface.co/descript/dac_16khz}} \cite{kumar2024high} offers a universal approach to audio compression, achieving an impressive 90x compression rate at 8 kbps for 44.1 kHz audio. It is designed to handle a wide range of audio signals while maintaining high fidelity. Speech Tokenizer\footnote{\url{https://github.com/ZhangXInFD/SpeechTokenizer.git}} \cite{zhang2024speechtokenizer} is a unified tokenizer for speech language models (SLMs) that employs RVQ to generate hierarchical representations capturing both linguistic and acoustic features. It demonstrates speech reconstruction quality comparable to EnCodec. EnCodec\footnote{\url{https://huggingface.co/facebook/encodec_24khz}} \cite{defossez2022high} is a high-fidelity NAC that features a streaming encoder-decoder architecture combined with RVQ for efficient audio compression. It is designed to preserve fine-grained audio details while achieving effective compression. All input audios are resampled to 16 kHz before being processed by DAC, Soundstream, and Speech Tokenizer, while EnCodec processes audio at 24 kHz. We extract CBRs from the frozen encoders of these codecs using average pooling, resulting in feature dimensions of 256 for Soundstream, 251 for DAC, 250 for Speech Tokenizer, and 375 for EnCodec.

\section{Modeling}

In this section, we detail the modeling approaches with individual RLRs and CBRs as well as the proposed framework, \textbf{\texttt{HYFuse}} for fusion of heterogenous RLRs and CBRs. For modeling individual representations, we use fully connected network (FCN) and CNN. The CNN has two 1D convolutional layers with 64 and 128 filters and a kernel size of 3. ReLU is the activation function used in the convolutional layers. The output is then flattened and passed through a FCN block with a dense layer of 128 neurons, followed by output ayer for classification which uses softmax as activation function. For the FCN model, we use same the modeling as the FCN block in the CNN. \par

\subsection{HYFuse}
We propose, \textbf{\texttt{HYFuse}} for the effective fusion of RLRs and CBRs. The architecture is presented in Figure \ref{archi}. \textbf{\texttt{HYFuse}} leverages hyperbolic geometry to fuse CBRs and RLRs while preserving their hierarchical relationships. The fusion in hyperbolic space complements both fine-grained acoustic details and high-level prosodic structures from the RLRs and CBRs respectively, resulting in more expressive and structured feature representations. Unlike euclidean fusion methods, which may distort the intrinsic organization of representations, hyperbolic fusion maintains relative distances and ensures that complementary features are optimally integrated. Detailed walkthrough of \textbf{\texttt{HYFuse}} is given as follows. The representations are first passed through 1D convolutional layers with the same architecture as used for modeling individual RLRs and CBRs. The features are then flattened and ready to be transformed to hyberbolic space. The transformation from euclidean to hyperbolic space is achieved using the exponential map:
\begin{equation}
\exp_0(x) =
\begin{cases} 
\text{tanh}(\kappa \lVert x \rVert) \frac{x}{\lVert x \rVert}, & \text{if } \lVert x \rVert > 0, \\
0, & \text{if } \lVert x \rVert = 0,
\end{cases}
\end{equation}
\noindent where $\kappa > 0$ denotes the curvature of the hyperbolic space, and $\lVert x \rVert$ represents the Euclidean norm of input feature. Now, the transformed features of RLRs and CBRs will be represented as $x_1$ and $y_1$. They are then fused through the Möbius addition operation. Möbius addition between the two hyperbolic points $x_1$ and $y_1$ is given by:
\begin{equation}
x_1 \oplus y_1 = \frac{(1 + 2 \langle x_1, y_1 \rangle + \lVert y_1 \rVert^2)x_1 + (1 - \lVert x_1 \rVert^2)y_1}{1 + 2 \langle x_1, y_1 \rangle + \lVert x_1 \rVert^2 \lVert y_1 \rVert^2},
\end{equation}
where $\langle x_1, y_1 \rangle$ denotes the Euclidean dot product, and $\lVert \cdot \rVert^2$ represents the squared Euclidean norm. Once fused, the resultant representation $y$ is mapped back to Euclidean space using the logarithmic map:
\begin{equation}
\log_0(y) = 
\begin{cases} 
2 \cdot \text{arctanh}(\lVert y \rVert) \frac{y}{\lVert y \rVert}, & \text{if } \lVert y \rVert < 1, \\
0, & \text{if } \lVert y \rVert = 0.
\end{cases}
\end{equation}
Ensuring that $\lVert y \rVert < 1$ maintains numerical stability within the Poincaré ball. FCN block with a dense layer is attached on top of the final fused representation followed by the output layer with softmax activation which outputs probabilities of the emotion classes. \textbf{\texttt{HYFuse}} trainable parameters for different input representations are from 8 to 13 millions.

\begin{table}[!bt]
\setlength{\tabcolsep}{4pt}
\scriptsize
\centering
\begin{tabular}{lcccccccc}
\toprule
\multicolumn{1}{c|}{\multirow{3}{*}{\textbf{Rep}}} & \multicolumn{4}{c|}{\textbf{CREMA-D}} & \multicolumn{4}{c}{\textbf{Emo-DB}} \\ 
\cmidrule(lr){2-5} \cmidrule(lr){6-9} 
\multicolumn{1}{c|}{} & \multicolumn{2}{c|}{\textbf{FCN}} & \multicolumn{2}{c|}{\textbf{CNN}} & \multicolumn{2}{c|}{\textbf{FCN}} & \multicolumn{2}{c}{\textbf{CNN}} \\ 
\cmidrule(lr){2-3} \cmidrule(lr){4-5} \cmidrule(lr){6-7} \cmidrule(lr){8-9}
\multicolumn{1}{c|}{} & \textbf{Acc} & \multicolumn{1}{c|}{\textbf{F1}} & \textbf{Acc} & \multicolumn{1}{c|}{\textbf{F1}} & \textbf{Acc} & \multicolumn{1}{c|}{\textbf{F1}} & \textbf{Acc} & \textbf{F1} \\ 
\midrule
\multicolumn{9}{c}{\textbf{RLRs}} \\ 
\midrule
\multicolumn{1}{l|}{W2} & \cellcolor{strawberry5}58.66 & \multicolumn{1}{c|}{\cellcolor{strawberry5}55.14} & \cellcolor{strawberry5}65.16 & \multicolumn{1}{c|}{\cellcolor{strawberry5}65.08} & \cellcolor{strawberry15}88.21 & \multicolumn{1}{c|}{\cellcolor{strawberry15}86.42} & \cellcolor{strawberry15}\textbf{91.51} & \cellcolor{strawberry15}\textbf{90.65} \\
\multicolumn{1}{l|}{W}  & \cellcolor{strawberry10}64.71 & \multicolumn{1}{c|}{\cellcolor{strawberry10}61.63} & \cellcolor{strawberry10}68.81 & \multicolumn{1}{c|}{\cellcolor{strawberry5}68.64}  & \cellcolor{strawberry10}87.23 & \multicolumn{1}{c|}{\cellcolor{strawberry5}85.21}  & \cellcolor{strawberry10}89.72 & \cellcolor{strawberry10}89.52 \\
\multicolumn{1}{l|}{XE} & \cellcolor{strawberry5}63.69 & \multicolumn{1}{c|}{\cellcolor{strawberry5}61.20} & \cellcolor{strawberry5}68.77 & \multicolumn{1}{c|}{\cellcolor{strawberry10}68.67} & \cellcolor{strawberry5}85.65 & \multicolumn{1}{c|}{\cellcolor{strawberry5}85.25}  & \cellcolor{strawberry5}81.31 & \cellcolor{strawberry5}80.61 \\
\multicolumn{1}{l|}{H}  & \cellcolor{strawberry15}67.85 & \multicolumn{1}{c|}{\cellcolor{strawberry15}66.25} & \cellcolor{strawberry15}70.63 & \multicolumn{1}{c|}{\cellcolor{strawberry15}70.45} & \cellcolor{strawberry5}86.25 & \multicolumn{1}{c|}{\cellcolor{strawberry10}85.39}  & \cellcolor{strawberry5}88.81 & \cellcolor{strawberry5}87.85 \\ 
\midrule
\multicolumn{9}{c}{\textbf{CBRs}} \\ 
\midrule
\multicolumn{1}{l|}{E}  & \cellcolor{strawberry10}47.85 & \multicolumn{1}{c|}{\cellcolor{strawberry10}46.98} & \cellcolor{strawberry5}48.48 & \multicolumn{1}{c|}{\cellcolor{strawberry5}42.56}  & \cellcolor{strawberry10}47.65 & \multicolumn{1}{c|}{\cellcolor{strawberry10}45.08} & \cellcolor{strawberry5}48.04 & \cellcolor{strawberry10}40.20 \\
\multicolumn{1}{l|}{D}  & \cellcolor{strawberry5}42.52 & \multicolumn{1}{c|}{\cellcolor{strawberry5}41.86} & \cellcolor{strawberry5}43.84 & \multicolumn{1}{c|}{\cellcolor{strawberry5}35.74}  & \cellcolor{strawberry5}39.78 & \multicolumn{1}{c|}{\cellcolor{strawberry5}38.52} & \cellcolor{strawberry5}40.56 & \cellcolor{strawberry5}34.10 \\
\multicolumn{1}{l|}{ST} & \cellcolor{strawberry5}41.61 & \multicolumn{1}{c|}{\cellcolor{strawberry5}40.14} & \cellcolor{strawberry10}48.51 & \multicolumn{1}{c|}{\cellcolor{strawberry10}45.92} & \cellcolor{strawberry5}45.65 & \multicolumn{1}{c|}{\cellcolor{strawberry5}43.71}  & \cellcolor{strawberry10}49.91 & \cellcolor{strawberry5}37.20 \\
\multicolumn{1}{l|}{SS} & \cellcolor{strawberry15}54.20 & \multicolumn{1}{c|}{\cellcolor{strawberry15}53.94} & \cellcolor{strawberry15}55.19 & \multicolumn{1}{c|}{\cellcolor{strawberry15}53.03} & \cellcolor{strawberry15}59.12 & \multicolumn{1}{c|}{\cellcolor{strawberry15}57.96} & \cellcolor{strawberry15}\textbf{61.36} & \cellcolor{strawberry15}\textbf{58.96} \\ 
\bottomrule
\end{tabular}
\caption{Performance Evaluation of models trained with various RLRs and CBRs; Scores are in \% and average of five folds; Acc and F1 stands for accuracy and marco-average F1 score; The 
abbreviations given are: Wav2vec2 (W2), WavLM (W), x-vector (XE), HuBERT (H), EnCodec (E), DAC (D), Speech Tokenizer (ST), Soundstream (SS); The abbreviations used herre are kept same for Table \ref{tab:fusion}}
\label{tab:single}
\end{table}

\section{Experiments}

\subsection{Dataset}

\noindent \textbf{CREMA-D} \cite{cao2014crema}: It contains 7,442 samples from 91 actors, representing a diverse range of racial and ethnic backgrounds, including Caucasian, African American, Hispanic, and Asian participants. The dataset features 48 male and 43 female actors, aged between 20 and 74 (average age: 36), offering a broad demographic coverage. Each actor delivers 12 sentences and spans over six emotions—happy, sad, anger, fear, disgust, and neutral. \newline
\noindent \textbf{Emo-DB} \cite{burkhardt2005database}: It comprises approximately 800 utterances recorded by 10 actors (5 male, 5 female), each performing a set of 10 carefully selected sentences expressing seven emotions: neutral, anger, fear, joy, sad, disgust, and boredom. It is a german SER corpus. % The recordings underwent rigorous perception testing, with only those achieving over 80\% recognizability and 60\% naturalness included in the final dataset. This ensures high-quality and reliable emotional speech data for research purposes.
\noindent Due to differences in audio duration, the NACs will produce different length representations. So as a initial preprocessing step, we pad the audios to the length of the maximum duration audio in the respective dataset for all our experiments. \newline
\noindent\textbf{Training Details}: The models are trained using cross-entropy as the loss function and Adam as the optimizer. We set the batch size as 32, learning rate as 1e-5, and epochs as 50. We leverage dropout and early stopping to mitigate overfitting. We follow 5-fold cross-validation for training and validating our models where 4 folds are used as training set and 1 fold as test set. 

\subsection{Experimental Results}
We present the evaluation of downstream models trained with individual RLRs and CBRs in Table \ref{tab:single}. We see that RLRs outperform CBRs across both datasets, indicating that CBRs struggle to capture the speech characteristics necessary for better SER. This performance is also observed across previous research evaluating RLRs and CBRs for SER \cite{mousavi2024dasb}. Among CBRs, Soundstream shows the strongest performance in both the datasets. However, even the best-performing CBRs still lags behind the RLRs. Among RLRs, HuBERT report that top performance with CNN in CREMA-D and Wav2vec2 with CNN in Emo-DB. This mixed performance points towards the effect of downstream data distribution on the performance of the models trained with representations. Overall, we see that CNN based models shows better performance than FCN models. These scores will be considered as baselines for experiments with combinations of different representations. \par
Table \ref{tab:fusion} presents the results of homogeneous (RLRs + RLRs, CBRs + CBRs) and heterogeneous (RLRs + CBRs) fusions of representations. We use concatenation (Concat) based fusion as the baseline fusion technique. We follow the same architecture as \texttt{\textbf{HYFuse}} up to feature flattening and subsequently applying a FCN with the same modeling details as \texttt{\textbf{HYFuse}}. We also keep the training details same as \texttt{\textbf{HYFuse}} for fair comparison. Our findings reveal that \texttt{\textbf{HYFuse}} consistently outperforms both individual representations and concatenation-based fusion across CREMA-D and Emo-DB, reinforcing the strength of hyperbolic transformation in aligning representations. When examining homogeneous fusion, we observe that while combining two RLRs, such as Wav2vec2 and HuBERT, yields improvements over its individual performances. This improvements in performances after combinations shows some emergence of complementary behavior in the representations. Also, fusing CBRs with CBRs yields lower performance than fusion of RLRs with RLRs and this is due to inherently lower individual performance of CBRs. However, we observe a surprising criterion as the fusion of some high performing RLRs and CBRs, shows far better performance than the homogeneous fusion of RLRs and CBRs, despite the performance of individual CBRs and fusion of CBRs with CBRs is quite low. Such fusion leads to improved performance by leveraging the complementary strengths of RLRs and CBRs, where RLRs provide rich prosodic information while CBRs capture fine-grained acoustic characteristics. \par

\begin{table}[!bt]
\scriptsize
\setlength{\tabcolsep}{4pt}
\centering
\begin{tabular}{lcccccccc}
\toprule
\multicolumn{1}{c|}{\multirow{3}{*}{\textbf{Pairs}}} & \multicolumn{4}{c|}{\textbf{CREMA-D}}                                                  & \multicolumn{4}{c}{\textbf{Emo-DB}}                                       \\ 
\cmidrule(lr){2-5} \cmidrule(lr){6-9} 
\multicolumn{1}{c|}{}                                      & \multicolumn{2}{c|}{\textbf{Concat}} & \multicolumn{2}{c|}{\textbf{\texttt{HYFuse}}}             & \multicolumn{2}{c|}{\textbf{Concat}} & \multicolumn{2}{c}{\textbf{\texttt{HYFuse}}} \\ 
\cmidrule(lr){2-3} \cmidrule(lr){4-5} \cmidrule(lr){6-7} \cmidrule(lr){8-9}
\multicolumn{1}{c|}{}                                      & \textbf{Acc}      & \textbf{F1}      & \textbf{Acc}      & \multicolumn{1}{c|}{\textbf{F1}}      & \textbf{Acc}      & \textbf{F1}      & \textbf{Acc}      & \textbf{F1}     \\ 
\midrule
\multicolumn{9}{c}{\textbf{RLRs + RLRs}} \\ 
\midrule
\multicolumn{1}{l|}{W2 + W}    & \cellcolor{strawberry5}60.98 & \cellcolor{strawberry5}59.65 & \cellcolor{strawberry5}64.58 & \multicolumn{1}{c|}{\cellcolor{strawberry5}63.38} & \cellcolor{strawberry10}88.63 & \cellcolor{strawberry5}87.36 & \cellcolor{strawberry5}93.36 & \cellcolor{strawberry5}92.27 \\
\multicolumn{1}{l|}{W2 + XE}   & \cellcolor{strawberry5}58.78 & \cellcolor{strawberry5}57.73 & \cellcolor{strawberry5}66.88 & \multicolumn{1}{c|}{\cellcolor{strawberry5}65.53} & \cellcolor{strawberry5}87.69 & \cellcolor{strawberry5}86.64 & \cellcolor{strawberry5}91.45 & \cellcolor{strawberry5}90.03 \\
\multicolumn{1}{l|}{W2 + H}    & \cellcolor{strawberry15}71.18 & \cellcolor{strawberry15}70.36 & \cellcolor{strawberry10}76.61 & \multicolumn{1}{c|}{\cellcolor{strawberry10}75.52} & \cellcolor{strawberry5}87.25 & \cellcolor{strawberry5}86.13 & \cellcolor{strawberry15}94.63 & \cellcolor{strawberry15}94.48 \\
\multicolumn{1}{l|}{W + XE}    & \cellcolor{strawberry10}69.33 & \cellcolor{strawberry10}68.18 & \cellcolor{strawberry5}74.49 & \multicolumn{1}{c|}{\cellcolor{strawberry5}73.38} & \cellcolor{strawberry5}88.24 & \cellcolor{strawberry5}87.34 & \cellcolor{strawberry5}92.08 & \cellcolor{strawberry5}91.36 \\
\multicolumn{1}{l|}{W + H}     & \cellcolor{strawberry5}65.97 & \cellcolor{strawberry5}65.82 & \cellcolor{strawberry15}77.25 & \multicolumn{1}{c|}{\cellcolor{strawberry15}76.69} & \cellcolor{strawberry15}89.64 & \cellcolor{strawberry15}88.61 & \cellcolor{strawberry10}94.25 & \cellcolor{strawberry5}93.37 \\
\multicolumn{1}{l|}{XE + H}    & \cellcolor{strawberry5}68.52 & \cellcolor{strawberry5}67.79 & \cellcolor{strawberry5}73.64 & \multicolumn{1}{c|}{\cellcolor{strawberry5}72.28} & \cellcolor{strawberry5}88.33 & \cellcolor{strawberry10}87.79 & \cellcolor{strawberry5}93.62 & \cellcolor{strawberry10}93.57 \\
\midrule
\multicolumn{9}{c}{\textbf{CBRs + CBRs}} \\ 
\midrule
\multicolumn{1}{l|}{E + D}     & \cellcolor{strawberry5}58.97 & \cellcolor{strawberry5}47.61 & \cellcolor{strawberry5}64.68 & \multicolumn{1}{c|}{\cellcolor{strawberry5}63.26} & \cellcolor{strawberry15}58.96 & \cellcolor{strawberry5}52.45 & \cellcolor{strawberry5}64.85 & \cellcolor{strawberry5}63.28 \\
\multicolumn{1}{l|}{E + ST}    & \cellcolor{strawberry5}58.97 & \cellcolor{strawberry5}45.08 & \cellcolor{strawberry10}66.67 & \multicolumn{1}{c|}{\cellcolor{strawberry10}65.59} & \cellcolor{strawberry10}58.68 & \cellcolor{strawberry15}56.51 & \cellcolor{strawberry5}64.14 & \cellcolor{strawberry5}63.68 \\
\multicolumn{1}{l|}{E + SS}    & \cellcolor{strawberry5}57.10 & \cellcolor{strawberry5}40.07 & \cellcolor{strawberry5}66.48 & \multicolumn{1}{c|}{\cellcolor{strawberry5}65.52} & \cellcolor{strawberry5}55.69 & \cellcolor{strawberry5}52.38 & \cellcolor{strawberry15}68.73 & \cellcolor{strawberry15}67.78 \\
\multicolumn{1}{l|}{D + ST}    & \cellcolor{strawberry5}55.23 & \cellcolor{strawberry5}51.44 & \cellcolor{strawberry5}63.43 & \multicolumn{1}{c|}{\cellcolor{strawberry5}62.52} & \cellcolor{strawberry5}55.96 & \cellcolor{strawberry10}54.09 & \cellcolor{strawberry5}61.19 & \cellcolor{strawberry5}60.05 \\
\multicolumn{1}{l|}{D + SS}    & \cellcolor{strawberry15}61.78 & \cellcolor{strawberry10}60.06 & \cellcolor{strawberry5}66.64 & \multicolumn{1}{c|}{\cellcolor{strawberry5}65.53} & \cellcolor{strawberry5}52.85 & \cellcolor{strawberry5}41.98 & \cellcolor{strawberry5}65.06 & \cellcolor{strawberry5}58.54 \\
\multicolumn{1}{l|}{ST + SS}   & \cellcolor{strawberry5}59.63 & \cellcolor{strawberry5}58.46 & \cellcolor{strawberry15}67.64 & \multicolumn{1}{c|}{\cellcolor{strawberry15}66.68} & \cellcolor{strawberry5}58.61 & \cellcolor{strawberry5}53.28 & \cellcolor{strawberry10}66.62 & \cellcolor{strawberry10}65.53 \\
\midrule
\multicolumn{9}{c}{\textbf{RLRs + CBRs}} \\ 
\midrule
\multicolumn{1}{l|}{W2 + E}    & \cellcolor{strawberry5}75.13 & \cellcolor{strawberry5}75.07 & \cellcolor{strawberry5}60.24 & \multicolumn{1}{c|}{\cellcolor{strawberry5}60.58} & \cellcolor{strawberry5}78.55 & \cellcolor{strawberry5}78.50 & \cellcolor{strawberry5}90.77 & \cellcolor{strawberry5}89.72 \\
\multicolumn{1}{l|}{W2 + D}    & \cellcolor{strawberry15}77.70 & \cellcolor{strawberry15}77.70 & \cellcolor{strawberry10}78.26 & \multicolumn{1}{c|}{\cellcolor{strawberry10}78.19} & \cellcolor{strawberry5}85.33 & \cellcolor{strawberry5}85.05 & \cellcolor{strawberry5}89.77 & \cellcolor{strawberry5}89.72 \\
\multicolumn{1}{l|}{W2 + ST}   & \cellcolor{strawberry5}72.41 & \cellcolor{strawberry5}71.89 & \cellcolor{strawberry5}74.62 & \multicolumn{1}{c|}{\cellcolor{strawberry5}74.39} & \cellcolor{strawberry5}84.15 & \cellcolor{strawberry5}83.79 & \cellcolor{strawberry5}91.84 & \cellcolor{strawberry5}91.52 \\
\multicolumn{1}{l|}{\textbf{W2 + SS}}   & \cellcolor{strawberry10}76.15 & \cellcolor{strawberry10}76.15 & \cellcolor{strawberry15}\textbf{79.29} & \multicolumn{1}{c|}{\cellcolor{strawberry15}\textbf{79.10}} & \cellcolor{strawberry5}82.57 & \cellcolor{strawberry5}82.24 & \cellcolor{strawberry15}\textbf{95.33} & \cellcolor{strawberry15}\textbf{95.18} \\
\multicolumn{1}{l|}{W + E}     & \cellcolor{strawberry5}66.89 & \cellcolor{strawberry5}66.79 & \cellcolor{strawberry15}77.68 & \multicolumn{1}{c|}{\cellcolor{strawberry15}76.82} & \cellcolor{strawberry5}80.37 & \cellcolor{strawberry5}79.49 & \cellcolor{strawberry15}95.20 & \cellcolor{strawberry15}94.11 \\
\multicolumn{1}{l|}{W + D}     & \cellcolor{strawberry5}60.91 & \cellcolor{strawberry5}60.55 & \cellcolor{strawberry5}66.13 & \multicolumn{1}{c|}{\cellcolor{strawberry5}65.75} & \cellcolor{strawberry5}85.05 & \cellcolor{strawberry5}84.71 & \cellcolor{strawberry5}85.98 & \cellcolor{strawberry5}85.98 \\
\multicolumn{1}{l|}{W + ST}    & \cellcolor{strawberry5}76.02 & \cellcolor{strawberry5}72.61 & \cellcolor{strawberry5}75.35 & \multicolumn{1}{c|}{\cellcolor{strawberry5}75.14} & \cellcolor{strawberry5}83.18 & \cellcolor{strawberry5}83.02 & \cellcolor{strawberry10}95.05 & \cellcolor{strawberry10}95.05 \\
\multicolumn{1}{l|}{W + SS}    & \cellcolor{strawberry5}65.33 & \cellcolor{strawberry5}65.28 & \cellcolor{strawberry5}66.89 & \multicolumn{1}{c|}{\cellcolor{strawberry5}66.99} & \cellcolor{strawberry5}64.96 & \cellcolor{strawberry5}63.01 & \cellcolor{strawberry5}87.31 & \cellcolor{strawberry5}85.98 \\
\multicolumn{1}{l|}{XE + E}    & \cellcolor{strawberry5}63.64 & \cellcolor{strawberry5}63.47 & \cellcolor{strawberry5}67.85 & \multicolumn{1}{c|}{\cellcolor{strawberry5}66.28} & \cellcolor{strawberry10}86.03 & \cellcolor{strawberry10}85.98 & \cellcolor{strawberry5}94.39 & \cellcolor{strawberry5}94.14 \\
\multicolumn{1}{l|}{XE + D}    & \cellcolor{strawberry5}65.16 & \cellcolor{strawberry5}65.01 & \cellcolor{strawberry5}69.63 & \multicolumn{1}{c|}{\cellcolor{strawberry5}68.32} & \cellcolor{strawberry5}86.02 & \cellcolor{strawberry5}85.05 & \cellcolor{strawberry5}91.59 & \cellcolor{strawberry5}91.59 \\
\multicolumn{1}{l|}{XE + ST}   & \cellcolor{strawberry5}65.28 & \cellcolor{strawberry5}65.14 & \cellcolor{strawberry5}71.52 & \multicolumn{1}{c|}{\cellcolor{strawberry5}70.89} & \cellcolor{strawberry15}87.30 & \cellcolor{strawberry15}86.92 & \cellcolor{strawberry5}92.52 & \cellcolor{strawberry5}92.20 \\
\multicolumn{1}{l|}{XE + SS}   & \cellcolor{strawberry5}63.65 & \cellcolor{strawberry5}63.53 & \cellcolor{strawberry5}69.88 & \multicolumn{1}{c|}{\cellcolor{strawberry5}68.13} & \cellcolor{strawberry5}69.31 & \cellcolor{strawberry5}65.14 & \cellcolor{strawberry5}93.01 & \cellcolor{strawberry5}92.52 \\
\multicolumn{1}{l|}{H + E}     & \cellcolor{strawberry5}67.02 & \cellcolor{strawberry5}66.51 & \cellcolor{strawberry5}69.18 & \multicolumn{1}{c|}{\cellcolor{strawberry5}69.04} & \cellcolor{strawberry5}85.56 & \cellcolor{strawberry5}85.05 & \cellcolor{strawberry5}88.35 & \cellcolor{strawberry5}87.85 \\
\multicolumn{1}{l|}{H + D}     & \cellcolor{strawberry5}64.14 & \cellcolor{strawberry5}63.99 & \cellcolor{strawberry5}68.23 & \multicolumn{1}{c|}{\cellcolor{strawberry5}68.47} & \cellcolor{strawberry5}82.62 & \cellcolor{strawberry5}82.24 & \cellcolor{strawberry5}90.14 & \cellcolor{strawberry5}88.79 \\
\multicolumn{1}{l|}{H + ST}    & \cellcolor{strawberry5}68.03 & \cellcolor{strawberry5}68.03 & \cellcolor{strawberry5}67.29 & \multicolumn{1}{c|}{\cellcolor{strawberry5}67.22} & \cellcolor{strawberry5}83.18 & \cellcolor{strawberry5}82.92 & \cellcolor{strawberry5}86.17 & \cellcolor{strawberry5}85.98 \\
\multicolumn{1}{l|}{H + SS}    & \cellcolor{strawberry5}67.02 & \cellcolor{strawberry5}66.82 & \cellcolor{strawberry5}68.83 & \multicolumn{1}{c|}{\cellcolor{strawberry5}68.64} & \cellcolor{strawberry5}65.96 & \cellcolor{strawberry5}63.21 & \cellcolor{strawberry5}89.31 & \cellcolor{strawberry5}88.79 \\ 
\bottomrule
\end{tabular}
\caption{Performance evaluation of model trained on combination of various RLRs and CBRs: The scores are presented in \% and average of five-folds}
\label{tab:fusion}
\end{table}

\begin{figure}[!bt]
    \centering
    \subfloat[]{%
        \includegraphics[width=0.23\textwidth]{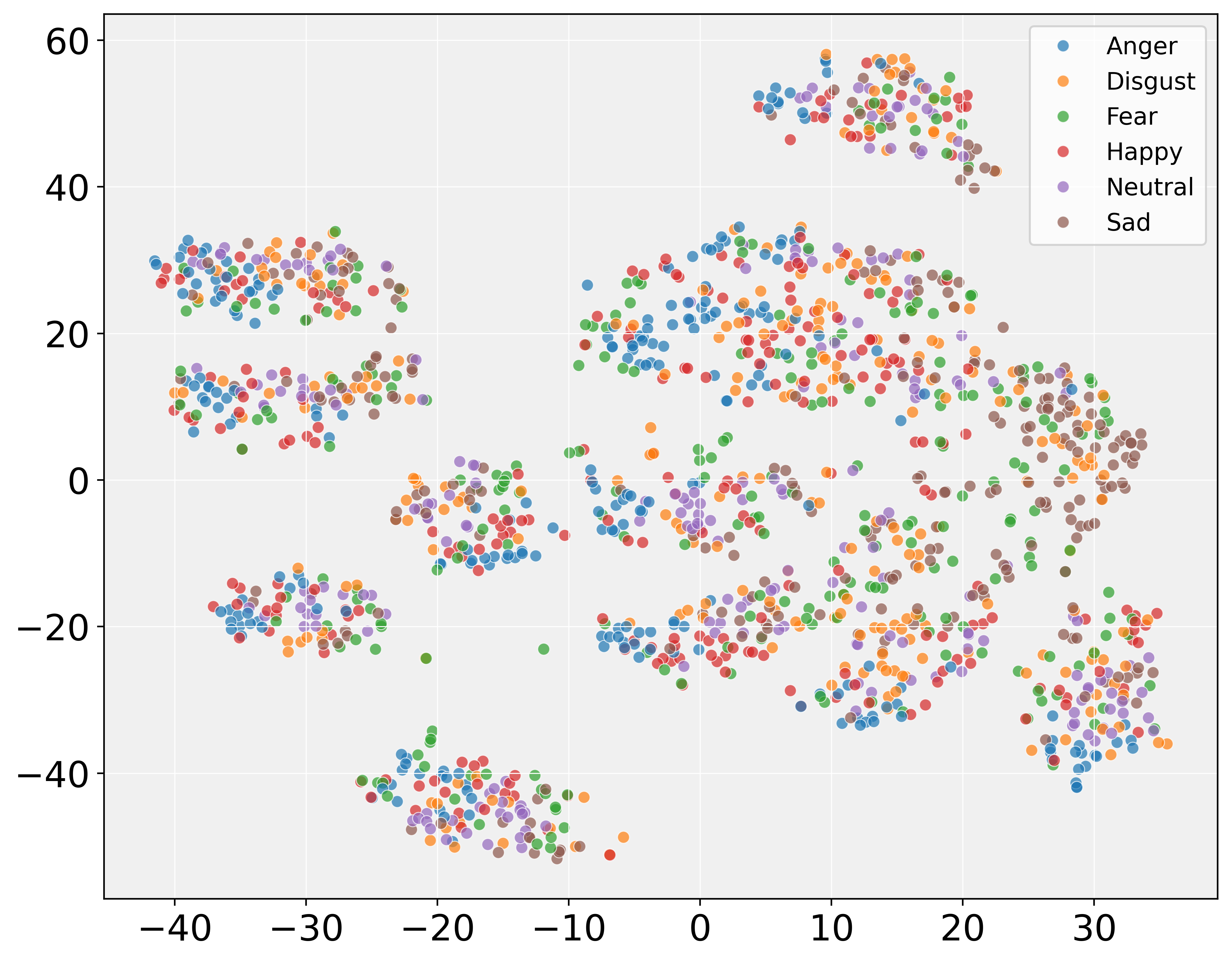}
    }
    \hfill
    \subfloat[]{%
        \includegraphics[width=0.23\textwidth]{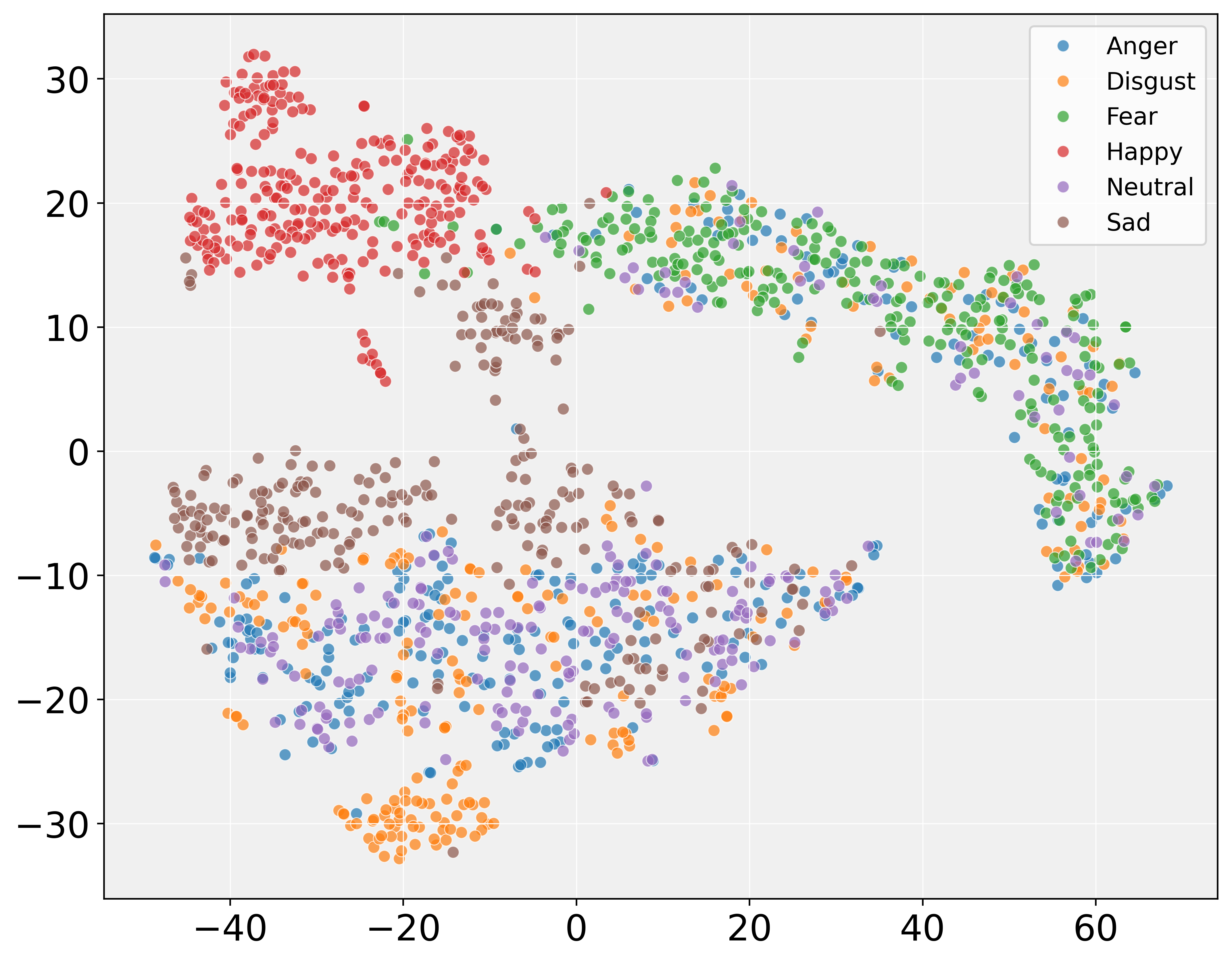}
    }
   
    \caption{t-SNE visualizations: (a) CNN (HuBERT) (b) \textbf{\texttt{HYFuse (Wav2vec2 + Soundstream)}}}
    \label{fig:tsne}
\end{figure}
% \vspace{-0.3cm}
\begin{figure}[!bt]
    \centering
    \subfloat[]{%
        \includegraphics[width=0.21\textwidth]{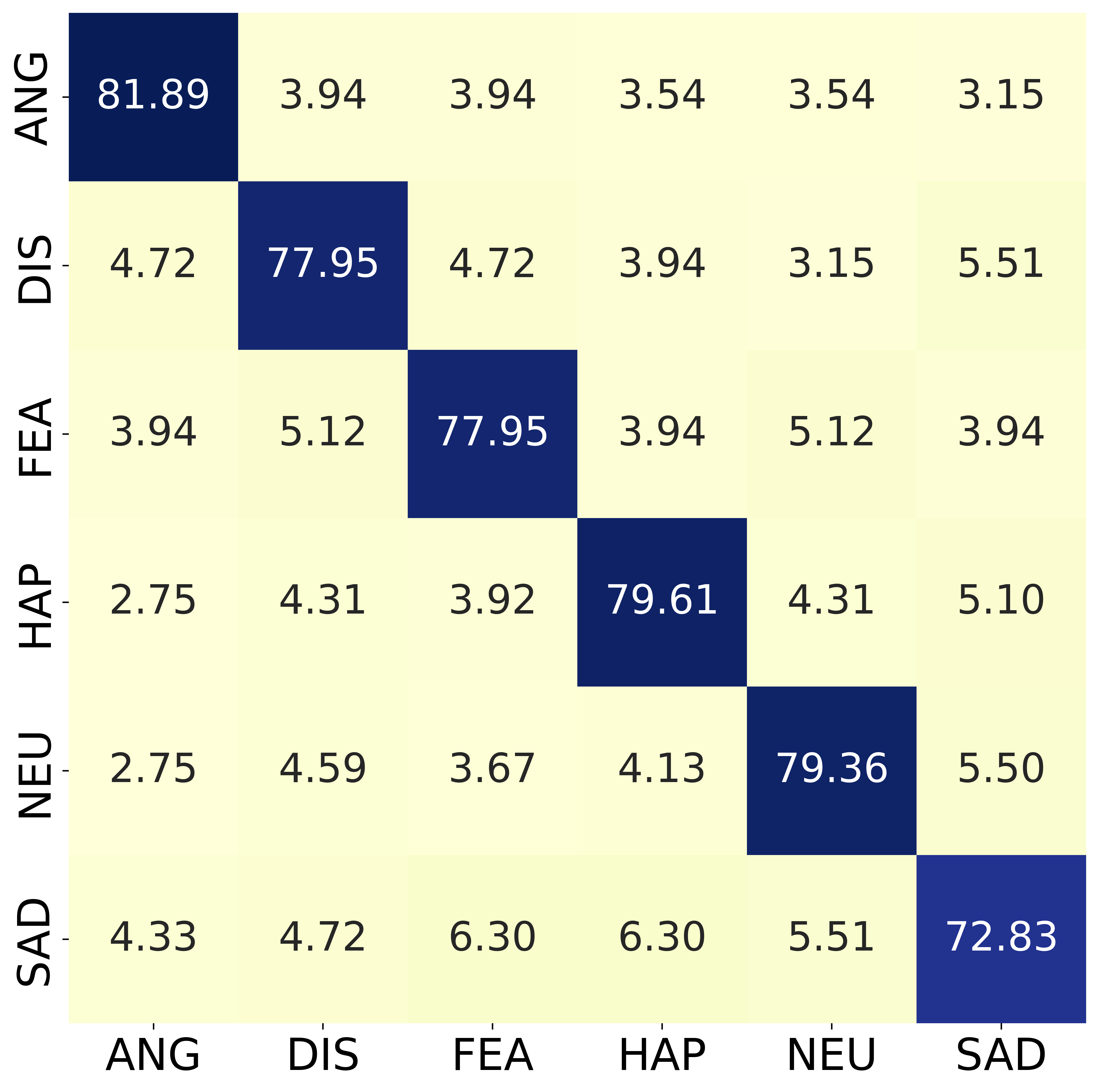}
    }
    \hspace{0.1cm}
    \subfloat[]{%
        \includegraphics[width=0.21\textwidth]{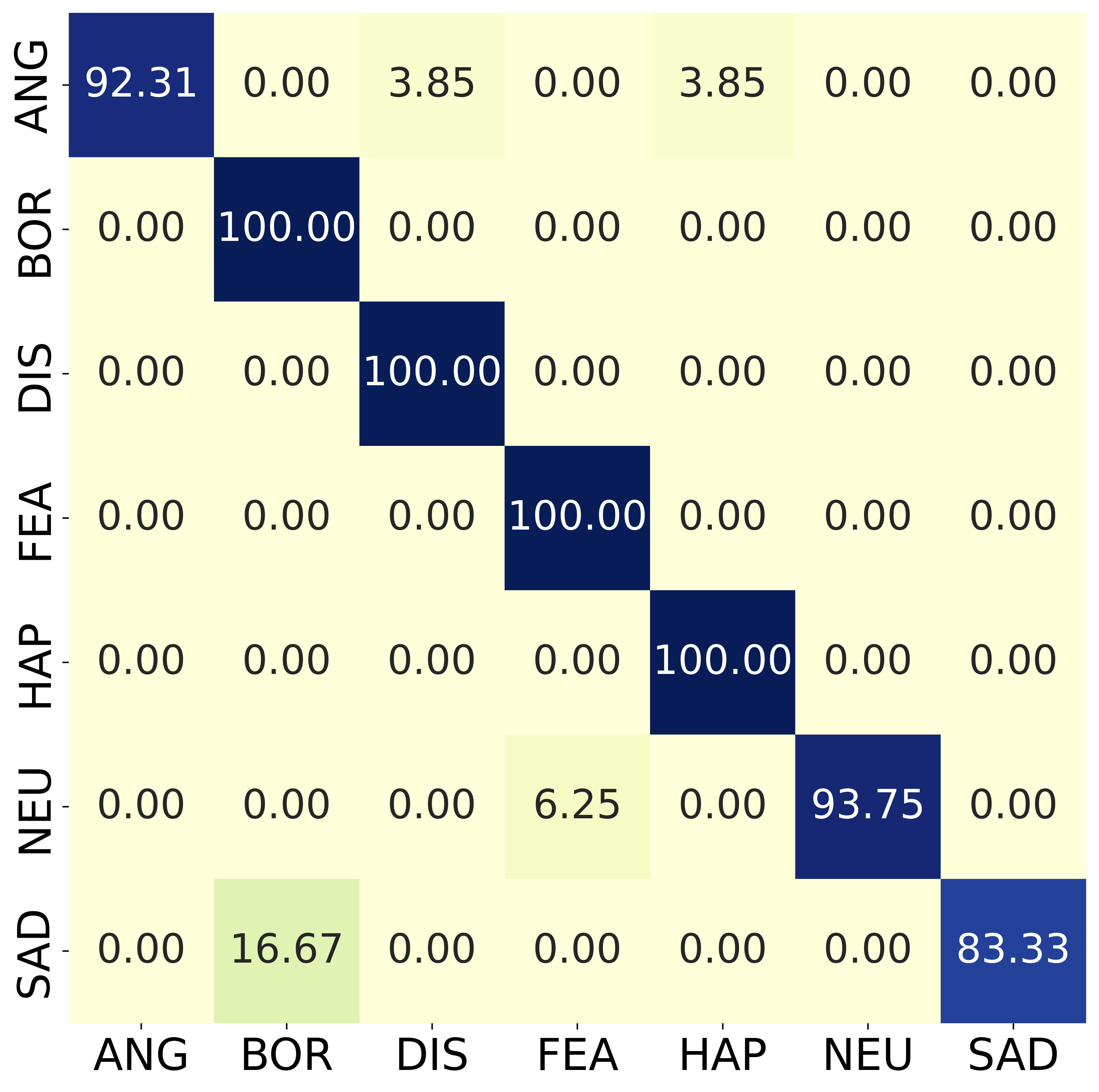}
    }
    \caption{Confusion matrices: (a) CREMA-D, (b) Emo-DB; y-axis indicates the True Values, whereas the x-axis represents the Predicted Values.}
    \label{fig:cm}
\end{figure}

\noindent This behavior is observed across both baseline concatenation-based fusion technique and fusion with \textbf{\texttt{HYFuse}}. However, the fusion through \textbf{\texttt{HYFuse}} brings out the complementary behavior more effectively by aligning the representations in hyperbolic space, thereby preserving their hierarchical relationships and minimizing distortion. Notably, the fusion of Wav2vec2 and Soundstream with \textbf{\texttt{HYFuse}} emerges as the best-performing combination, surpassing all individual representations, homogeneous representations fusion, and the baseline concatenation-based fusion technique. With this performance, we achieve SOTA performance as the individual RLRs were SOTA for SER \cite{yang21c_interspeech}. Further, this highlights the advantage of integrating heterogeneous representations for improved. This validates \textit{our hypothesis that heterogeneous fusion of RLRs and CBRs will be most effective for SER.} Additionally, not all the combinations of RLRs with CBRs leads to improved performance over homogeneous fusion of RLRs, however, we believe that fusion of strong individual CBRs with strong individual RLRs brings out the best in them. Aside from the best-performing combination of Wav2vec2 and Soundstream, we observe that the fusion of WavLM and EnCodec also shows competitive performance, though it does not surpass the top combination. This suggests that while the integration of complementary representations enhances performance, the degree of improvement depends on the specific characteristics of the fused representations. Overall, our findings highlight the significance of selecting the right combination of RLRs and CBRs, as well as the effectiveness of \textbf{\texttt{HYFuse}} in leveraging their strengths to further enhance SER. We present the t-SNE plot visualization of downstream trained on HuBERT (best performing representations for CREMA-D) and \textbf{\texttt{HYFuse}} with the fusion of Wav2vec2 and Soundstream in Figure \ref{fig:tsne} for CREMA-D. We extract the representations from the penultimate layer of the downstream models. We observe far better clustering of the emotion classes with \textbf{\texttt{HYFuse}}, thus showing its effectiveness for better SER. Additionally, we also plot the confusion matrices of \textbf{\texttt{HYFuse}} with Wav2vec2 and Soundstream for both the datasets in Figure \ref{fig:cm}.  \par

% \vspace{-0.5cm}
\section{Conclusion}
In this study, we explored the fusion of RLRs and CBRs for SER, addressing a previously unexplored research gap. While CBRs such as EnCodec capture intricate acoustic features like pitch and timbre, RLRs from pre-trained models like WavLM encode high-level prosodic information. We hypothesized that their fusion would enhance SER performance by leveraging their complementary strengths. To this end, we proposed \textbf{\texttt{HYFuse}}, a novel framework that transforms RLRs and CBRs into hyperbolic space for effective fusion. Through the integration of x-vector (RLR) and SoundStream (CBR), \textbf{\texttt{HYFuse}} outperforms individual representations and homogeneous fusion approaches, achieving SOTA performance. These findings highlight the potential of heterogeneous representation fusion in advancing SER and also as a reference for future research exploring such heterogeneous fusion.

\bibliographystyle{IEEEtran}
\bibliography{main}

\end{document}